\PassOptionsToPackage{unicode}{hyperref}
\PassOptionsToPackage{hyphens}{url}
\PassOptionsToPackage{dvipsnames,svgnames,x11names}{xcolor}
\documentclass[
]{article}
\usepackage{amsmath,amssymb}
\usepackage{lmodern}
\usepackage{iftex}
\ifPDFTeX
  \usepackage[T1]{fontenc}
  \usepackage[utf8]{inputenc}
  \usepackage{textcomp} 
\else 
  \usepackage{unicode-math}
  \defaultfontfeatures{Scale=MatchLowercase}
  \defaultfontfeatures[\rmfamily]{Ligatures=TeX,Scale=1}
\fi
\IfFileExists{upquote.sty}{\usepackage{upquote}}{}
\IfFileExists{microtype.sty}{
  \usepackage[]{microtype}
  \UseMicrotypeSet[protrusion]{basicmath} 
}{}
\makeatletter
\@ifundefined{KOMAClassName}{
  \IfFileExists{parskip.sty}{%
    \usepackage{parskip}
  }{
    \setlength{\parindent}{0pt}
    \setlength{\parskip}{6pt plus 2pt minus 1pt}}
}{
  \KOMAoptions{parskip=half}}
\makeatother
\usepackage{xcolor}
\usepackage{graphicx}
\makeatletter
\def\maxwidth{\ifdim\Gin@nat@width>\linewidth\linewidth\else\Gin@nat@width\fi}
\def\maxheight{\ifdim\Gin@nat@height>\textheight\textheight\else\Gin@nat@height\fi}
\makeatother
\setkeys{Gin}{width=\maxwidth,height=\maxheight,keepaspectratio}
\providecommand{\pandocbounded}[1]{#1}
\makeatletter
\def\fps@figure{htbp}
\makeatother
\NewDocumentCommand\citeproctext{}{}
\NewDocumentCommand\citeproc{mm}{%
  \begingroup\def\citeproctext{#2}\cite{#1}\endgroup}
\makeatletter
 \let\@cite@ofmt\@firstofone
 \def\@biblabel#1{}
 \def\@cite#1#2{{#1\if@tempswa , #2\fi}}
\makeatother
\newlength{\cslhangindent}
\newlength{\csllabelwidth}
\newenvironment{CSLReferences}[2] 
 {\begin{list}{}{%
  \setlength{\itemindent}{0pt}
  \setlength{\leftmargin}{0pt}
  \setlength{\parsep}{0pt}
  \ifodd #1
   \setlength{\leftmargin}{\cslhangindent}
   \setlength{\itemindent}{-1\cslhangindent}
  \fi
  \setlength{\itemsep}{#2\baselineskip}}}
 {\end{list}}
\usepackage{calc}

\ifLuaTeX
\usepackage[bidi=basic]{babel}
\else
\usepackage[bidi=default]{babel}
\fi
\babelprovide[main,import]{american}

\def\languageshorthands#1{}
\ifLuaTeX
  \usepackage{selnolig}  
\fi
\IfFileExists{bookmark.sty}{\usepackage{bookmark}}{\usepackage{hyperref}}
\IfFileExists{xurl.sty}{\usepackage{xurl}}{} 
\hypersetup{
  pdftitle={astrodatapy: a Python compilation of observational
constraints for galaxy formation and reionization models},
  pdfauthor={Yuxiang Qin, Manodeep Sinha},
  pdflang={en-US},
  colorlinks=true,
  linkcolor={Maroon},
  filecolor={Maroon},
  citecolor={Blue},
  urlcolor={Blue},
  pdfcreator={LaTeX via pandoc}}

\title{astrodatapy: a Python compilation of observational constraints
for galaxy formation and reionization models}

\definecolor{c53baa1}{RGB}{83,186,161}
\definecolor{c202826}{RGB}{32,40,38}

\usepackage[affil-it]{authblk}
\usepackage{orcidlink}
\author[1%
  ]{Yuxiang Qin%
    \,\orcidlink{0000-0002-4314-1810}\,%
    }
\author[2%
  ]{Manodeep Sinha%
    \,\orcidlink{0000-0002-4845-1228}\,%
    }

\affil[1]{Research School of Astronomy and Astrophysics, Australian
National University, Canberra, ACT 2611, Australia%
  }
\affil[2]{Australia's Climate Simulator (ACCESS-NRI), Australian
National University, Canberra, ACT, Australia%
  }
\date{30 November 2026}

\begin{document}
\maketitle

\section{Summary}\label{summary}

Models of galaxy formation and cosmic reionization are calibrated and
tested against statistical measurements of the galaxy and quasar
populations and of the intergalactic medium (IGM): how many galaxies
there are per unit volume at a given stellar mass or luminosity, how
their properties correlate, and how the ionization state and temperature
of the IGM evolve with redshift. \texttt{astrodatapy} is a Python
package that collects these measurements in one place. Version 2.4.0
holds 307 datasets from 187 papers in 29 quantities: stellar mass, star
formation rate, ultraviolet (UV) and {[}O III{]} luminosity functions of
galaxies; quasar luminosity functions in the UV, optical, X-ray and
bolometric luminosity and black hole mass functions; scaling relations
such as the black hole--bulge, Tully--Fisher, size--mass, gas
fraction--mass and stellar-to-halo mass relations; quasar clustering;
and the evolution of the IGM ionizing emissivity, H I photoionization
rate, mean free path of ionizing photons, temperature and neutral
fraction. A single call loads every dataset of a quantity near a target
redshift and returns it converted to the Hubble parameter, initial mass
function (IMF) and units the user asks for (\autoref{fig:example}).

\section{Statement of need}\label{statement-of-need}

The measurements a model is compared with are published in heterogeneous
forms. Number densities are given per Mpc\(^3\) or per \(h^{-3}\)
Mpc\(^3\) for different values of the Hubble parameter \(h\)
(\citeproc{ref-Croton2013}{Croton, 2013}); stellar masses and star
formation rates assume a Salpeter (\citeproc{ref-Salpeter1955}{1955}),
Kroupa (\citeproc{ref-Kroupa2001}{2001}) or Chabrier
(\citeproc{ref-Chabrier2003}{2003}) IMF; luminosity functions use
different magnitude systems and bands; quantities are tabulated in
linear or logarithmic form with symmetric, asymmetric or logarithmic
errors; some points are upper or lower limits; lengths are comoving or
proper; some results are only given as fitted Schechter or power-law
parameters, and some only as figures. Every modelling group re-collects
and re-converts these data, conversions are a common source of silent
errors, and the exact data behind a published comparison are rarely
recorded.

\texttt{astrodatapy} stores each dataset once, in the units and
conventions of its paper, with those conventions recorded as metadata,
and performs the conversions when the data are loaded. It is aimed at
modellers who calibrate semi-analytic, semi-numerical or hydrodynamical
models of galaxy formation and reionization and at observers who place a
new measurement among previous ones. Because every dataset carries a
bibcode and a header that cites the table or figure its values come
from, a comparison made with a given \texttt{astrodatapy} version can be
reproduced and traced back to the literature.

\section{State of the field}\label{state-of-the-field}

Several simulation codes ship the observational data they are compared
with: the semi-analytic models Shark (\citeproc{ref-Lagos2018}{Lagos et
al., 2018}), Galacticus (\citeproc{ref-Benson2012}{Benson, 2012}) and
L-Galaxies (\citeproc{ref-Henriques2015}{Henriques et al., 2015}), and
the empirical model UniverseMachine
(\citeproc{ref-Behroozi2019}{Behroozi et al., 2019}). These collections
serve one code: the data are converted to that code's assumptions in
advance, or the corrections for \(h\) and the IMF are written into the
script that reads each dataset, and they are not distributed as
libraries for other codes. Inference frameworks such as 21CMMC
(\citeproc{ref-Greig2015}{Greig \& Mesinger, 2015};
\citeproc{ref-Park2019}{Park et al., 2019}) bundle the few datasets
their likelihoods use. Compilations of a single observable collect one
kind of measurement, such as the quasar luminosity functions homogenised
by Shen et al. (\citeproc{ref-Shen2020}{2020}) or the published upper
limits on the 21-cm power spectrum from reionization and cosmic dawn in
\texttt{eor\_limits} (\citeproc{ref-eorlimits}{EoRImaging, 2026}), which
provides them as human-readable files with plotting tools. The closest
general-purpose tool is CoReCon (\citeproc{ref-Garaldi2023}{Garaldi,
2023}), a community collection of reionization constraints.
\texttt{astrodatapy} aims to handle the data more directly: it keeps
each measurement as published, records its conventions, and converts it
to the user's conventions when it is loaded.

\section{Software design}\label{software-design}

\begin{figure}
\centering
\pandocbounded{\includegraphics[keepaspectratio]{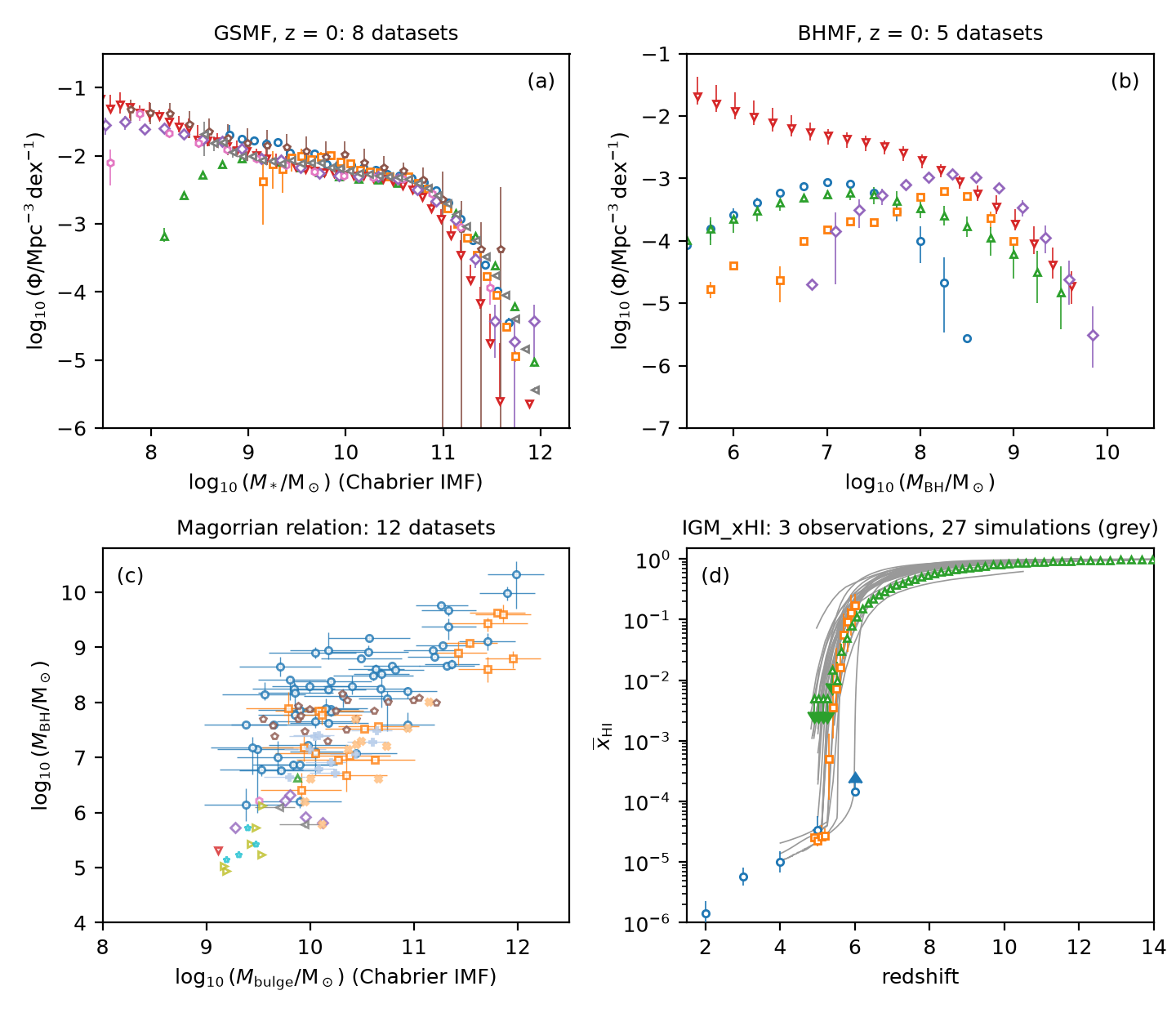}}
\caption{Four of the 29 quantities in \texttt{astrodatapy} 2.4.0, all
converted to \(h = 0.6751\): (a) the galaxy stellar mass function at
\(z = 0\) and (c) the black hole--bulge mass relation, with stellar
masses converted to a Chabrier (\citeproc{ref-Chabrier2003}{2003}) IMF;
(b) the black hole mass function at \(z = 0\); (d) the volume-averaged
neutral hydrogen fraction of the IGM against redshift, with simulation
curves in grey. Arrows mark upper and lower limits.\label{fig:example}}
\end{figure}

The package separates data from code. Each quantity (``feature'') is a
directory of plain-text files: an \texttt{info.txt} with one row per
dataset (data format, logarithmic or linear values, normalisation,
Hubble parameter and its power, IMF, comoving or proper units, dataset
type, measurement method and bibcode), and one data file per dataset and
redshift with a header that documents the source. Plain text was chosen
over a binary or database format so that every value can be read, diffed
and reviewed in a pull request, and so that contributors can add data
without writing code. Datasets keep their published values rather than
values pre-converted to one convention: the provenance stays checkable
against the paper, and any target convention can be requested
(\autoref{fig:example}).

Four loader classes cover the shapes of the data:
\texttt{number\_density} for distribution functions at a given redshift,
\texttt{correlation} for scaling relations, \texttt{clustering} for
correlation lengths and \texttt{evolution} for quantities tabulated
against redshift. Each returns a \texttt{pandas}
(\citeproc{ref-McKinney2010}{McKinney, 2010}) table of datasets whose
data are \texttt{numpy} (\citeproc{ref-Harris2020}{Harris et al., 2020})
arrays of values with upper and lower bounds, together with flags for
upper and lower limits, the dataset type and the method. The conversions
are applied in a fixed order (in \texttt{number\_density}: evaluation of
parametric fits, IMF, magnitude band, error format, limits, logarithm,
normalisation and Hubble parameter), so a new dataset needs only a row
of metadata. A limit is an ordinary row of its dataset, marked by its
error columns: \texttt{nan} for an upper limit and, for quantities
tabulated against redshift, \texttt{inf} for a lower limit. These
markers pass through every conversion unchanged, and the loader then
flags the rows as limits. Simulation curves are stored in the same
format and marked \texttt{data\_sim}, so they can be drawn next to the
measurements without being mistaken for them.

\begin{figure}
\centering
\pandocbounded{\includegraphics[keepaspectratio]{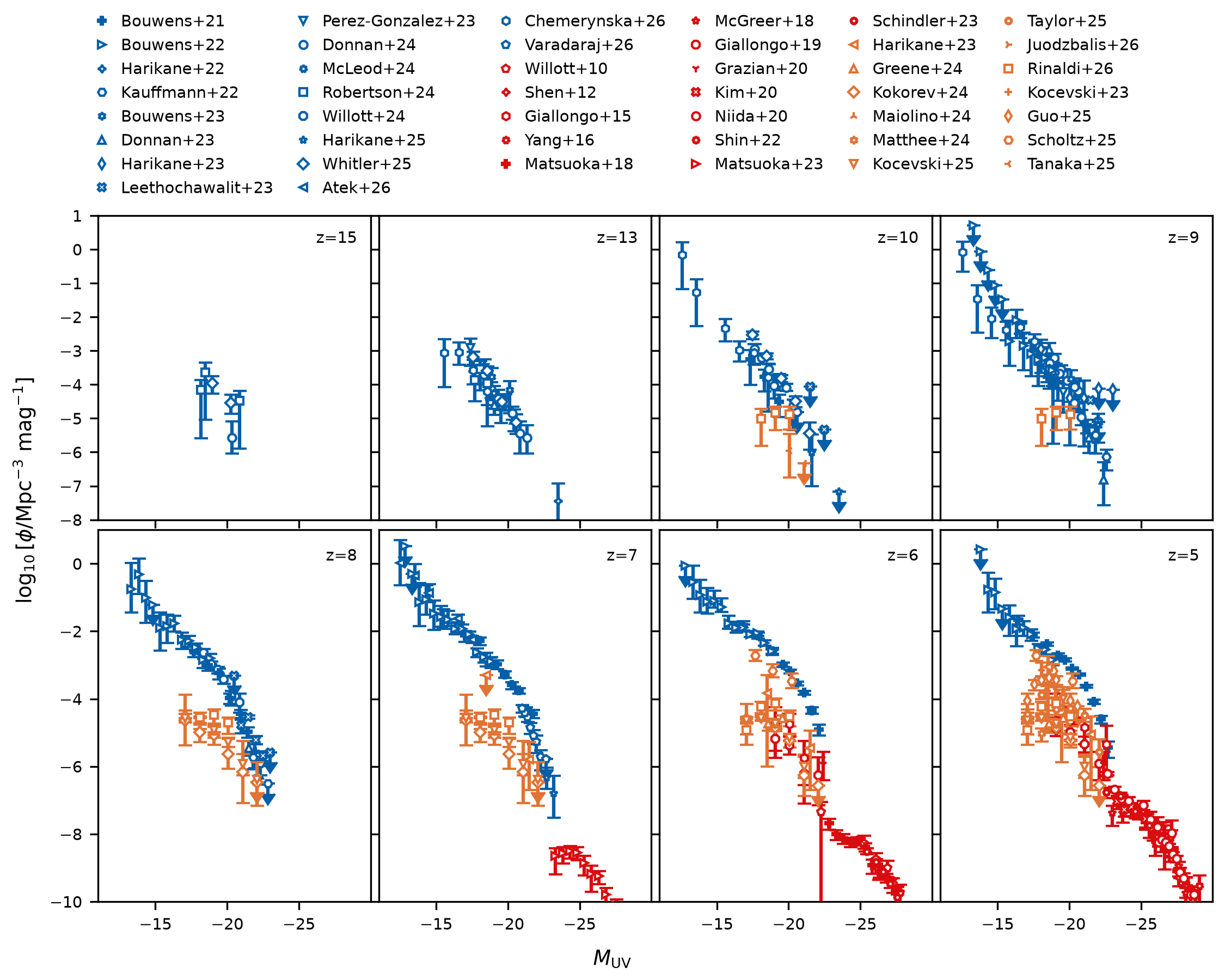}}
\caption{Rest-frame UV luminosity functions of galaxies (blue; 18
papers) and of quasars and AGN (red: conventional measurements; orange:
JWST-selected faint AGN and Little Red Dots; 26 papers in total) at
\(z = 5\)--15 in \texttt{astrodatapy} 2.4.0, converted to
\(h = 0.6751\), as drawn by the package's \texttt{combine.ipynb}
notebook without its combined constraints. Arrows mark upper
limits.\label{fig:uvlf}}
\end{figure}

The test suite loads every dataset of every feature, checks unit
conversions against hand-computed values, and executes the example
notebooks, which plot with Matplotlib (\citeproc{ref-Hunter2007}{Hunter,
2007}); it runs on Python 3.7 to 3.14 in continuous integration, and the
documentation, built on every change, executes the notebooks and
generates a catalogue of all datasets from the \texttt{info.txt} files.
One notebook combines the high-redshift galaxy and quasar UV and {[}O
III{]} luminosity functions into binned constraints, archived on Zenodo
(\citeproc{ref-Qin2026zenodo}{Qin, 2026}); \autoref{fig:uvlf} shows the
UV measurements it combines.

\section{Research impact statement}\label{research-impact-statement}

\texttt{astrodatapy} has been used since 2018 to calibrate and validate
the Meraxes semi-analytic model of the DRAGONS programme
(\citeproc{ref-Mutch2016}{Mutch et al., 2016}). The Meraxes analysis
package \texttt{dragons} requires it for its comparisons with the
observed stellar mass, star formation rate, UV luminosity, quasar
luminosity and black hole mass functions, and the Meraxes calibration
driver \texttt{mhysa} lists it as a dependency. Published studies that
used it include the Meraxes work on the sizes and morphologies of
high-redshift galaxies (\citeproc{ref-Marshall2019}{Marshall et al.,
2019}) and on the evolution of black hole scaling relations
(\citeproc{ref-Marshall2020}{Marshall et al., 2020}), whose public
analysis scripts load the comparison data with \texttt{astrodatapy}, and
the study of \(z \gtrsim 12\) galaxies observed by JWST of Qin et al.
(\citeproc{ref-Qin2023}{2023}), which used it to convert the
observations to a common IMF. Outside this collaboration, the stellar
mass function plotting tool of the MP-Gadget simulation code has used it
since 2019, and the Markov chain Monte Carlo calibration framework of
the SAGE semi-analytic model (\texttt{sage-mcmc}) takes its
observational constraints from it. Five contributors besides the authors
have added data or code through pull requests, and issues from users
have shaped the package; for example, the per-point bin widths it has
returned since version 2.1.0 were requested for likelihood analyses.

\section{AI usage disclosure}\label{ai-usage-disclosure}

The work of November 2026 that led to versions 1.0.0 to 2.4.0 was done
with the assistance of Claude Code (Anthropic), using the Claude Opus
5.5 model; every commit it co-authored carries a \texttt{Co-Authored-By}
trailer. It assisted with porting the package to Python 3 and modern
packaging, writing tests, continuous integration and documentation,
auditing existing datasets against their source papers, transcribing
tables and reading curves from published figures for new datasets, and
drafting this paper. YQ reviewed every change; from version 1.0.0
onwards each was proposed as a pull request and merged by YQ. New and
corrected values were checked against the source papers, and the scripts
and checks used to read values from figures are kept with the data
provenance. The design of the package and of its data format, and the
choice of datasets, are the authors'. The code and data written between
2018 and July 2026 were produced without generative AI.

\section{Acknowledgements}\label{acknowledgements}

We thank Balu Sreedhar, Fatma Shaban, Madeline Marshall, Simon Mutch and
Varun Nikam for their contributions to the package, and Emanuele Maria
Ventura for issues and discussions that improved it. We gratefully
acknowledge the HPC support from OzSTAR and Gadi in Australia. YQ is
supported by the Australian Research Council (ARC) Discovery Early
Career Researcher Award (DECRA) through fellowship \#DE240101129.

\section*{References}\label{references}
\addcontentsline{toc}{section}{References}

\protect\phantomsection\label{refs}

\end{document}